\documentclass[12pt]{iopart}
\usepackage{iopams}
\begin{document}
\title{Irreducible tensor-form of the relativistic corrections to the M1 transition operator}
\author{M. Nemouchi\dag \ and M.R. Godefroid\ddag}
\address{\dag\ Laboratoire d'\'Electronique Quantique, Facult\'e de Physique, USTHB, BP32,
El-Alia, Algiers, Algeria}
\address{\ddag\ Chimie Quantique et Photophysique, CP160/09, Universit\'e Libre de Bruxelles, \\ 
Av. F.D. Roosevelt 50, B-1050
Brussels, Belgium}
\ead{mrgodef@ulb.ac.be}
%%%%%%%%
\begin{abstract}
The relativistic corrections to the magnetic dipole moment operator in the Pauli approximation 
were derived originally by Drake~\cite{Dra:71a}. In the present paper, we derive their irreducible
 tensor-operator form to be used in atomic structure codes adopting the Fano-Racah-Wigner 
 algebra for calculating its matrix elements. 
%seminal
\end{abstract}
\pacs{31.10.+z,31.30.jc,32.70.Cs}

\submitto{\jpb}
\noindent{\it Keywords\/}:
magnetic dipole transitions, Breit-Pauli calculations, irreducible tensors, operator techniques

\vfill
\noindent{\hfill \bf \today}

\maketitle

\section{Introduction}

For a long time, theoretical studies of atomic transition probabilities have mostly concentrated 
on electric dipole (E1) allowed transitions, usually responsible for the strongest lines in atomic spectra. 
It has been quickly realized that forbidden transitions such as electric quadrupole (E2) and 
magnetic dipole (M1) transitions would gain in intensity in astrophysical and low-density laboratory 
plasmas where collisional deexcitation of metastable states is low enough to buildup the population of 
metastable states \cite{Gar:62b,Ost:89a}. 
 These forbidden lines are indeed observed in a wide range of astronomical objects such as planetary nebulae, Wolf–Rayet stars, novae and the Sun, and they are important in providing information on the electron temperature and density of stellar objects. In some situations, these lines can be used for calibrating spectrometers over wide wavelength ranges  \cite{Kucetal:00a}. 
It is now recognized that more than half of the atoms in the Universe 
recombined via forbidden channels, so that their accurate treatment is crucial in order to follow the 
cosmological recombination process with the level of precision required by future 
microwave anisotropy experiments \cite{WonSco:07a}. 

Amongst the forbidden processes, magnetic dipole transitions often played a key role. 
A review on M1 transitions in Hydrogen 
and H-like ions has been written by Sucher~\cite{Suc:78a}.
Helium-like systems occupy a special position in atomic physics as the simplest 
multielectron systems for testing the theoretical calculation of transition amplitudes \cite{Suletal:08a}. 
As illustrated by 
Johnson {\it et al.} in their review \cite{Johetal:95a}, a great deal of attention 
has also been given to the magnetic dipole transitions $2 ^3S_1 \rightarrow 1 ^1S_0$ along the helium isoelectronic 
sequence in the period 1970-1995. Historically, the observation of the spectral line associated to the 
transition $1s2s \; ^{3}S\longrightarrow 1s^{2} \;^{1}S$ in helium-like ions was explained 
by Breit and Teller \cite{BreTel:40a} as a two-photon electric dipole (2E1) process, 
the M1 transition amplitude being strictly zero when using the ``classical'' transition operator.
Later, Gabriel and Jordan \cite{GabJor:69a,GabJor:69b} identified some coronal solar lines as 
arising from the transition $1s2s \; ^{3}S\longrightarrow 1s^{2}\;^{1}S$ in helium-like 
C V, O VII, Ne IX, Mg XI and Si XIII,  
and proposed that $1s2s \; ^{3}S$ would decay to the ground state mostly through the M1 channel. 
Griem \cite{Gri:69a} demonstrated theoretically on the basis of relativistic calculations that 
this conjecture was founded.

But the interest of magnetic dipole 
transitions is not limited to one- and two-electron systems.
After Bowen's interpretation of the strongest nebular transitions~\cite{Bow:28a},  Condon \cite{Con:34a} found that the nebulium 
lines N$_1$ and N$_2$ of O~III  were essentially due to magnetic dipole radiation. The theoretical 
tools for estimating
M1 transition probabilities were then developped by Pasternack \cite{Pas:40a} and Shortley \cite{Sho:40a} in $LS$ coupling and in 
the intermediate coupling. The solar coronal lines 
remained a longstanding spectrocopy mystery until Edl\'{e}n \cite{Edl:42a} found the wave numbers 
coincidences with M1 transitions within the ground configurations $3s^23p$, $ 3s^23p^2$, $ 3s^23p^4$ and 
$ 3s^23p^5$ in Fe and Ni \cite{Swi:43a}. A bit later, M1 transitions were identified by 
Edl\'{e}n \cite{Edl:44a} in the configurations $5p^5$ and $5p^4$ and $6p^5$ of Xenon and Radon.
  Forbidden lines have now long been used in the analysis of solar, stellar or nebular astrophysical plasmas.  
  In the infrared, they are amongst the strongest stellar lines, and many of those of the light atoms have 
  been measured. Spectra of the solar corona and solar flares are rich in M1 lines in the visible and 
  ultraviolet   for the elements nitrogen through nickel. From the eighties, these lines have achieved 
  new importance for   diagnostics of low-density, magnetically-confined, high-temperature plasmas 
  generated in tokamaks, as   illustrated by the compilation of Kaufman and Sugar  \cite{KauSug:86a} 
  of the forbidden lines within $ns^2 np^k$   configurations. The majority of the reported observed 
  lines belonging to elements with $ Z \leq 20 $ were observed in astrophysical spectra, while the lines 
  belonging to elements with $ Z > 28 $ were observed in spectra of low density, magnetically confined, 
  high temperature laboratory plasmas. This compilation has been extended recently by Feldman and 
  Doschek \cite{FelDos:07a} including lines in the $500-1600$~\AA 
  ~range emitted by solar coronal plasmas 
  and recorded by the Solar Ultraviolet Measurements of Emitted Radiation (SUMER) spectrometer.   
Magnetic dipole transitions within the lowest $3d^4 \; ^5D$ term in the Ti I isoelectronic sequence, 
  violating   the general wavelength scaling behavior (ie moving rapidly to shorter wavelengths as Z increases)  and therefore suitable for plasma diagnostics in high-energy Tokamak fusion devices, have been found 
  by Serpa  \cite{Seretal:96a,Seretal:97a,Watetal:01a}. 
  
  The theory of magnetic dipole transitions in crystals has been often overlooked in the literature but
  their importance in the  description of the spectroscopic properties of lanthanides in crystals 
  has been stressed recently \cite{WybSme:07a}. \\

  In the next two subsections of the introduction,  we illustrate the increasing interest in M1/E2 transitions of both experimentalist and theoretician communities by providing a list, far to be exhaustive,  of relevant publications that appeared during the last few years. A commented bibliography for forbidden transitions, including M1 transitions, in atoms and ions can be found in Bi\'emont and Zeippen \cite{BieZei:96a} covering the period 1989-1995, a helpful complement to their previous bibliography on the subject \cite{BieZei:91a}.

\subsection{Experimental work on M1 transitions}

After the theoretical review work by Johnson {\it et al} \cite{Johetal:95a}, experimental data 
became available for various low-Z elements \cite{Schetal:94a,Creetal:98a,Traetal:99a}, as 
illustrated by Tr\"abert {\it et al.}~\cite{Traetal:03b} who explored the heavy-ion storage ring technique for investigating the metastable level  $1s2s \; ^3S_1$ of Li${^+}$ and Be$^{2+}$.
 Radiative lifetimes in the microsecond regime were  measured using the Electron Beam 
 Ion Trap (EBIT) facility, allowing the precise measurement of the lifetime of the 
 $1s2s \; ^3S_1$ metastable level in heliumlike C$^{4+}$ \cite{Schetal:94a}, N$^{5+}$,  
 O$^{6+}$ \cite{Creetal:98a} and S$^{14+}$ \cite{Creetal:06a}. This kind of experimental work is crucial for testing  the atomic physics theories and is not limited to two-electron systems.
 These experiments  provide important checks on the theoretical calculations of the rate of the transitions and are not limited to heliumlike systems. The precision of measurements of M1/E2 transitions between the fine-structure levels in highly charged ions has been continously increased. The lifetimes of the $P$ levels producing the coronal transitions of Fe XIV (Al-like) and Fe X  (Cl-like)  have been measured in an electrostatic Kingdon trap \cite{MoeChu:99a}, producing the metastable ions with an electron cyclotron resonance ion source.  The same technique \cite{Moeetal:98a} has been used to obtain the magnetic dipole transition rates  from measured lifetimes of levels of Be-like and B-like ions \cite{MoeChu:98a}, and to measure the  lifetimes of metastable levels in the ground term of Fe ions within $3s^2 3p^k$, ($k=1-5$)\cite{Moeetal:01a}.  Excited-configuration metastable level lifetimes have also been obtained in Cl-like Mn~IX and Fe~X \cite{Moeetal:00a}. A similar experimental setup has been adopted for the measurement of the metastable lifetime of $2s^2 2p^2 \; ^1S_0$ in O$^{+2}$~\cite{Smietal:04a}. 
  The electron beam ion trap technique~\cite{Tra:07a} and/or the heavy-ion storage ring measurements \cite{Tra:02a,Tra:06a} have been explored to investigate the E1-forbidden transition probabilities between the fine-structure levels of the ground configuration or to measure the E1-forbidden excited levels decay rates in
 Be-like \cite{Bacetal:98a,Traetal:00a,Traetal:01a}, 
 B-like  \cite{Seretal:98b,Traetal:00a,Traetal:01a,Traetal:02a,Traetal:03a,Lapetal:05a,Lapetal:06a}, 
 C-like  \cite{Traetal:00c,Traetal:05b},
 N-like  \cite{Traetal:00c},
 F-like \cite{Traetal:00a,Traetal:01a}, 
 Al-like \cite{Beietal:03a,Tra:04a,Breetal:07a,Traetal:09a}, 
 Si-like \cite{Seretal:98b,Traetal:98a,Traetal:00b,Traetal:02b,Traetal:98a,Traetal:09a},
 P-like \cite{Traetal:02b,Traetal:06a,Traetal:08a,Traetal:09a}, 
 S-like \cite{TraGwi:01a,Traetal:02b,Traetal:04b,Traetal:08a,Traetal:09a},
 Cl-like \cite{Traetal:02b,Traetal:04a,Tra:04a,Traetal:06a}, 
 Ti-like \cite{Seretal:97a,Uttetal:00a}
ions. 
The accuracy achieved in the EBIT measurement of the $1s^2 2s^2 2p \; ^2P^o_{3/2}$ level lifetime in boronlike Argon \cite{Lapetal:05a} is such that it becomes sensitive to the relativistic electron correlation like the frequency-dependent Breit interaction, as well as the electron anomalous magnetic moment. 
From the lifetimes of $3s^2 3p^k$ ground configuration levels of Al-, Si-, P- and S-like ions of Fe, Co and Ni measured at a heavy-ion storage ring, Tr\"{a}bert {\it et al} \cite{Traetal:09a} observed that the  decay curves show strong evidence of cascade repopulation from specific 3d levels, implying that
the clean monoexponential (or few-exponential) decay signals observed in a number of earlier measurements at heavy-ion storage rings are not generally encountered.  
Atomic level lifetimes have been measured for magnetic dipole transitions in Si-like, P-like, Ar-like,
K-like, and Ca-like ions of Kr, using an electron-beam ion trap in magnetic trapping mode \cite{Traetal:01b}. 
The storage ring technique has been used for measuring radiative lifetimes of metastable levels in more complex atomic systems such as Eu~II~\cite{Rosetal:01a},  Fe~II~\cite{Rosetal:01b,Haretal:03a},  La~II~\cite{Deretal:02a}, Ti~II~\cite{Haretal:03b}, Ar~II~\cite{Shetal:04a}, Sc~II~\cite{Lunetal:08a} 
,  for which the selectivity in populating the closely spaced metastable levels is achieved using the laser probing technique \cite{Man:03a}.

\subsection{Theoretical calculations}

%MCHF+BP /CIV3
Einstein rate coefficients have been calculated  for magnetic dipole, electric quadrupole and magnetic quadrupole transitions in 	Be-like ions \cite{KinHib:01a}, using configuration interaction wavefunctions in the
Breit-Pauli approximation, completing previous studies covering selected ions \cite{Fleetal:95a,Fleetal:96b}. 
A simple expression for calculating the line strengths of the magnetic dipole transitions has been derived by Kingston and Hibbert \cite{KinHib:01b} and tested in Be-like ions. More recently, Breit-Pauli energy levels, lifetimes, and transition probabilities obtained from  variational MCHF wave functions, have been compiled for the beryllium-like to neon-like sequences \cite{FroTac:04a}. Breit-Pauli transition probabilities were computed by Froese Fischer and Rubin \cite{FroRub:04a} for the intraconfiguration E2 and M1 transitions in Fe$^{+3}$, an ion that is expected to be a significant fraction of gaseous iron in many nebulae. Relativistic energy levels, lifetimes and transition probabilities have also been obtained for the sodium-like to argon-like sequences \cite{Froetal:06a}, also using the full relativistic multiconfiguration Dirac-Hartree-Fock (MCDHF) method to ascertain the reliability of the Breit-Pauli approximation at high-degrees of ionization.
Radiative rates for electric quadrupole  and magnetic dipole  transitions among the
lowest 32 levels belonging to the $3d^6$ configuration of doubly charged iron (Fe~III) were
calculated by Deb and Hibbert \cite{DebHib:09a} using $J$-dependent configuration interaction  wavefunctions calculated within CIV3. Ion storage ring lifetimes measurements of metastable levels in Sc~II were assisted by Breit-Pauli CI calculations \cite{Lunetal:08a}.

%MBPT+BP
The many-body perturbation theory (MBPT) was used
to account for the electron correlation, including relativistic corrections
 in the Breit-Pauli approximation for evaluating E2 and M1 transition rates along the N I isoelectronic sequence  \cite{Meretal:99a}.  As far as M1 transition amplitudes are concerned, for all these Breit-Pauli calculations   whatever the method used for obtaining the zero-order non-relativistic wave functions (variational MCHF, MBPT, CI, \ldots), the relativistic corrections to the M1 transition operator that should be taken into account according to Drake \cite{Dra:71a},  have been systematically omitted. There is a large number of theoretical works on M1 transition probabilities  using {\sc SUPERSTRUCTURE} \cite{EisZei:81a} and its extension {\sc AUTOSTRUCTURE} \cite{Bad:86a,Bad:97a}, also adopting the Breit-Pauli approximation but including the relativistic corrections in the transition operator, after the seminal study of Zeippen \cite{Zei:82a} on the subject. Some contributions are presented in section~2, dedicated to that issue. 
 
 In the full relativistic scheme, the relativistic many-body perturbation theory (MBPT), including the Breit interaction, was applied to estimate transition rates for all M1 transitions within $2l 2l'$  and for some $2l 3l' - 2l 3l'$ transitions in Be-like ions with nuclear charges ranging from $Z=4$ to $Z=100$ \cite{Safetal:99b}. Relativistic MBPT calculations have also been performed for multipole (E1, M1, E2, M2, E3, and M3) transition wavelengths and rates between $3l^{-1}4l'$ 	excited and ground states in nickel-like ions 
\cite{Safetal:06a}. This theory agrees with MCDHF calculations in lowest-order, includes all 
second-order correlation corrections, and includes corrections from negative energy states.
The $4s^24p \; ^2P_{1/2-3/2}$ transition rates were  calculated using a similar method for Ga-like ions with nuclear charge Z ranging from 31 to 100 \cite{Safetal:06b}. All these calculations provide theoretical benchmarks for comparison with experiment and theory.
Transition rates and line strengths are calculated for electric-multipole (E2 and E3) and magnetic-multipole (M1,M2 and M3) transitions between $3s^2 3p^6 3d^{10}$, $3s^2 3p^6 3d^9 4l$, $3s^2 3p^5 3d^{10}4l$, and $3s 3p^6 3d^{10} 4l$ states with $4l=4s, 4p, 4d$ and $4f$ in Ni-like ions with nuclear charges ranging from $Z=34-100$ \cite{Safetal:08a}. More recently, the same approach was adopted to study the M1 contributions to the lifetimes of the  $5d_{3/2}$ and $5d_{5/2}$ metastable levels of Ba$^+$ \cite{IskSaf:08a}. Motivated by the need of accurate theoretical and laboratory data on  high-energy radiative emission of highly ionized tungsten ions in EUV and Xray wavelength regions, Vilkas {\it et al}
\cite{Viletal:08b} used the relativistic multireference M\o ller-Plesset perturbation theory for calculating energy levels and transition probabilities in Ne-like xenon, tungsten, and uranium ions.
% and in Na-like to P-like Xe	ions \cite{Viletal:08a} (NO M1, only E1)

% GRASP
The variational multi-configuration Dirac-Fock  method has been used to study the 
$2s^2 2p \; ^2P_{3/2-1/2}$ M1 transitions of B-like and all the M1 transitions of 
Be-like argon~\cite{Donetal:01a}.
On the basis of a multiconfiguration Dirac-Fock method, a systematic study
has been carried out for the decay process of the $1s 2s^2 \; ^2S_{1/2}$ state of Li-like
ions to discuss the balance between the Auger decay channel, the two-electron one-photon (TEOP) and
M1 radiative decay modes~\cite{Donetal:06a}.
Radiative rates are calculated for  E1,M1,E2,M2 transitions  in Cl-like Fe~X \cite{AggKee:04a} and in Boron-like to F-like Kr ions  \cite{Aggetal:08a}, using the GRASP code.
A multiconfiguration Dirac-Fock method has been applied \cite{Quietal:07a} for computing the wavelengths and transition probabilities for lines in the X-ray spectra of the gallium-like ions from Yb XL to U~LXII, including the forbidden transitions (M1 and E2) within the ground configuration $4s^2 4p$.
The E2 and M1 transition probabilities within the ground configuration $3d^5$ of Fe$^{3+}$ have been calculated by Froese Fischer {\it et al} \cite{Froetal:08b}, adopting the Multiconfiguration Dirac-Hartree-Fock approach. 

 The magnetic-dipole transition probabilities between the fine-structure levels for
B-like and  Be-like ions were calculated by Tupitsyn {\it et al.} \cite{Tupetal:05a}, using the configuration interaction method in a Dirac-Fock-Sturm basis, getting a fair agreement with observation, except for B-like Ar$^{13+}$ for which the discrepancies cannot 
 be explained by the recoil corrections {\cite{Voletal:08a}.

\section{Relativistic corrections to the M1 operator}

From a theoretical point of view, M1 amplitudes played a crucial role in the development of theoretical
atomic physics. 	For example,  the role of the negative continuum has been illustrated in the full 
relativistic calculation  of M1 transitions in two-electron ions using the Multiconfiguration 
Dirac-Fock Method \cite{Ind:96a}. Large contributions of negative-energy states to forbidden 
magnetic-dipole transition amplitudes have been also found in relativistic MBPT calculations in beryllium-like ions  \cite{Safetal:99b} and alkali-metal atoms \cite{Savetal:99a}.
The  estimation of M1 transition amplitudes in the Breit-Pauli approximation also raises interesting 
questions. Starting from the Dirac-Breit Hamiltonian and the semi-classical radiation theory, Drake  
\cite{Dra:71a} derived the expression of the relativistic magnetic dipole transition operator 
to be used in the Breit-Pauli scheme. He estimated the contribution of this decay mode to the 
radiative lifetime of the metastable $1s2s \; ^{3}S$ level along the helium isoelectronic 
series in the range $Z=2-26$, in good agreement with observation.
The relativistic corrections to the magnetic dipole moment operator in the Pauli 
approximation were derived independently by Drake~\cite{Dra:71a} using semiclassical 
radiation theory and by Feinberg and Sucher \cite{FeiSuc:71a} using conventional 
quantum electrodynamics. In their work, the last authors criticized Drake's approach 
for including both the Breit operator and transverse photon in the same Hamiltonian. 
In response to these criticisms, Drake \cite{Dra:72a} derived theorems showing that 
the Breit interaction remains valid in the presence of radiation emission and that 
the $O(\alpha^2 Z^2)$ corrections to radiative-transition probabilities obtained by 
the semiclassical method always agree with the quantum-electrodynamic results. 
A few years later, Lin~\cite{Lin:77a}  obtained a Foldy-Woulthuysen transformation 
for use in the field theory of quantum electrodynamics and clarified the remaining 
ambiguity concerning the treatment of the $\vec{A} ^2$ term from the interaction 
Hamiltonian. A completely different approach based on relativistic many-body perturbation 
theory  was adopted by Johnson {\it et al.} \cite{Johetal:95a} and  
Derevianko {\it et al.}~\cite{Deretal:98b}. Some disagreement with nonrelativistic 
results based on the Breit Hamiltonian incited  Lach and Pachucki~\cite{LacPac:01a} 
to derive rigorously nonrelativistically forbidden single-photon transition rates 
between low-lying states of the helium atom within quantum electrodynamics. 
More recently, Pachucki \cite{Pac:03a} extended this study to $E1$, $M1$, $E2$ and $M2$ 
forbidden transitions in light atoms, also investigating the role of the anomalous 
magnetic moment of the electron in some forbidden transitions~\footnote{some important 
corrections to the original M1 transition current \cite{Pac:03a} appeared in \cite{Pac:04a}.}.

The relativistic corrections to the M1 transition operator derived by Drake \cite{Dra:71a}
were shown to be relevant for more complex systems of astrophysical interest. 
The ratio of line intensities of forbidden transitions 
 $^{2}D^{o}_{5/2}-\; ^{4}S^{o}_{3/2}$ et $^{2}D^{o}_{3/2}-\; ^{4}S^{o}_{3/2}$ 
 within the half-filled $p^3$ atomic configuration provides an 
 interesting tool for plasma diagnostic (see for example \cite{Monetal:90a,Bau:06a}). 
 In the high electron density limit, this ratio is entirely determined by 
the radiative transition probabilities and given by
\begin{equation*}
  r(\infty)=\frac{3}{2} \frac{A^{E2}(^{2}D^{o}_{5/2}\rightarrow{^{4}S}^{o}_{3/2})+
  A^{M1}(^{2}D^{o}_{5/2}\rightarrow  {^{4}S}^{o}_{3/2})}{A^{M1}(^{2}D^{o}_{3/2}
  \rightarrow  {^{4}S}^{o}_{3/2})+
  A^{E2}(^{2}D^{o}_{3/2}\rightarrow  {^{4}S}^{o}_{3/2})} \; .
\end{equation*}
To explain the disagreement between the theoretical and observed values of this ratio in 
highly dense plasmas for  atomic systems such as 
O II \cite{SeaOst:57a} and N I \cite{Dopetal:76a},  Zeippen~\cite{Zei:80a} suggested the 
importance of the relativistic corrections to the magnetic dipole transition operator, as originally derived by Drake \cite{Dra:71a} for two-electron systems. Introducing these corrections  in the code 
{\sc SUPERSTRUCTURE}, 
Eissner and Zeippen~\cite{EisZei:81a} and Zeippen~\cite{Zei:82a}  calculated the 
ratio [O II] $I(3729)/I(3726)$. This value is considered as the best ratio
value to be used for estimating electron densities from observational spectroscopy~\cite{Wanetal:04a}. 
Further calculations have been performed by Zeippen and collaborators for O~II~\cite{Zei:87a} and 
N-like ions \cite{ButZei:84a,Becetal:89a}, getting improved radiative transition probabilities for 
forbidden lines in closer agreement with MCHF/Breit-Pauli values \cite{GodFro:84a} but systematically illustrating the effect of 
the relativistic corrections to the M1 operator.
The low- and high-density limits of the forbidden fine structure line intensity ratio 
[O II] $I(3729)/I(3726)$ have been reinvestigated by Pradhan {\it et al} \cite{Praetal:06a} using the 
Breit-Pauli R-matrix (BPRM) method, in considerable disagreement with the elaborate calculations
of McLaughlin and Bell \cite{McLBell:98a} that Keenan et al (1999) \cite{Keeetal:99a} used to get
a problematic low-density limit of 1.5. Montenegro {\it et al} \cite{Monetal:06a} showed that this 
outstanding discrepancy between the observed line intensity ratios of [OII], and those calculated using 
the earlier results of McLaughlin and Bell (1998) are not due to relativistic effects, which are 
negligibly small. A fully relativistic study of forbidden transitions of O II appeared 
recently \cite{Cheetal:07a}, reporting a high-density ratio in good agreement with values deduced 
from the astronomical observation of planetary nebulae~\cite{CopWri:02a,Wanetal:04a}.
 Similarly, Storey and Zeippen \cite{StoZei:00a} demonstrated the necessity of including the 
relativistic corrections to the magnetic dipole operator for an accurate estimation of the 
 M1 rates for $^1D_2 - \; ^3P_2$ and $^1D_2 - \; ^3P_1$ transitions  within the
 $2p^2$ and $2p^4$ ground configurations of carbon-like and oxygen-like ions, respectively.
  The average value for the ratio of observed fluxes of the [O III] $\lambda \lambda$ 4959, 5007 
  lines in spectra of active galaxies and quasars \cite{Dimetal:07a} 
supports the theoretical improvement by Storey and Zeippen \cite{StoZei:00a}.

The importance of the relativistic corrections to the M1 operator was also found in astronomical 
X-ray spectroscopy. The iron K lines that appear in emission in many natural X-ray sources \cite{Bauetal:03a},   
have a well-known plasma diagnostics potential. In the calculation of atomic data for K-vacancy states in 
Fe XXIV used in the spectral modeling of iron K lines \cite{Bauetal:03a}, the situation becomes critical 
for the $1s2s2p \; ^4P^o_{5/2}$ metastable level which is shown to decay 
through both M1 and M2 transitions. In their work, also based on the Breit-Pauli approximation, 
the authors observed that the M1 $A$-values must be calculated with the relativistically corrected operator, 
the difference with the uncorrected version reaching five orders of magnitude. \\

From a pragmatic point of view,  
the inclusion of the relativistic corrections to the M1 operator established by Drake \cite{Dra:71a} in the Breit-Pauli relativistic treatment of complex atoms, 
is currently limited to {\sc SUPERSTRUCTURE} \cite{Eisetal:74a} and 
to its extension, {\sc AUTOSTRUCTURE}   \cite{Bad:86a,Bad:97a}, using the Slater 
determinant approach. Other well known atomic structure codes allowing a relativistic treatment in
 the Breit-Pauli approach do exist ({\sc CIV3} \cite{Hib:75a}, R-matrix codes \cite{Beretal:95a}, 
 {\sc MCHF} \cite{Fro:91aa,Fro:00a}, {\sc ATSP2K} \cite{Froetal:07a} )  but systematically adopt the 
 non-relativistic version of the M1 transition operator \cite{Con:34a,Sob:72a,Cow:81a}. 
 These atomic structure codes use the Wigner-Fano-Racah algebra 
\cite{Fanetal:63a,Fan:65a} and its extensions \cite{Rud:97a} for evaluating the angular integration of 
the Hamiltonian and transition operators 
matrix elements \cite{Froetal:91a,FroGod:91a}. The starting point of this algebra is to find for each 
operator its expression in terms of irreducible spherical tensorial operators \cite{Jud:98a}. 
In the present work, we derive it for the relativistic corrections to the M1 
operator introduced by Drake  \cite{Dra:71a}. \\

\newpage

\section{The magnetic dipole transition operator}

The magnetic dipole (M1) decay rate (in s$^{-1}$) is given by~\cite{Cow:81a}
\begin{equation}
\label{eq:M1_decay_rate}
  A^{M1}(u \rightarrow l) = \frac{4}{3}  \frac{1}{(2J_u +1)} \frac{1}{\hbar } 
  \left( \frac{\omega}{c} \right) ^3
  \vert \langle \gamma_l J_l \Vert Q^{(1)} \Vert \gamma_u J_u \rangle \vert ^2
\end{equation}
where $u$ and $l$ refer to the upper and lower levels, respectively, $\omega$ is 
the angular frequency corresponding to the transition energy
$\Delta E_{ul} = \hbar \omega$  and $\langle J_l \Vert Q^{(1)}  \Vert J_u \rangle$ 
is the reduced matrix element of the magnetic dipole moment tensorial operator. 
The latter is built from the vectorial magnetic dipole moment components of
\begin{eqnarray}
\label{eq:M1_operator_1}
\fl \bi {Q}=\mu_{B}  \sum_{i=1}^{N}\Bigg\{(\bi {l}_{i}+2\bi
{s}_{i})\left(1-\frac{\bi
{p}_{i}^{2}}{2m^{2}c^{2}}-\frac{1}{10}\frac{\omega^{2}}{c^{2}}r_{i}^{2}\right)
+\frac{\hbar\omega}{2mc^{2}}\;\bi
{s}_{i}+\;\frac{1}{2m^{2}c^{2}}\;\bi {p}_{i}\wedge (\bi
{p}_{i}\wedge \bi
{s}_{i})\nonumber\\
\lo+
\left(\frac{1}{5}\frac{\omega^{2}}{c^{2}}-\frac{Ze^{2}}{mc^{2}r_{i}^{3}}\right)
\;\bi {r}_{i}\wedge (\bi {r}_{i}\wedge \bi {s}_{i}) +
\;\frac{e^{2}}{2mc^{2}r_{ij}}\sum_{i\neq j}\Bigg[\frac{\bi
{r}_{ij}\wedge [\bi {r}_{ij}\wedge (\bi {s}_{i}+\bi
{s}_{j})]}{r_{ij}^{2}}\nonumber\\
\lo+\frac{1}{2}\frac{(\bi {r}_{i}\wedge \bi {r}_{j})\bi
{r}_{ij}\cdot (\bi {p}_{i}+\bi {p}_{j})}{r_{ij}^{2}} -\frac{\bi
{r}_{i}\wedge \bi {p}_{j}+\bi {r}_{j}\wedge \bi
{p}_{i}}{2}\Bigg]\Bigg\}
\end{eqnarray}
expressed in the Gaussian (mixed) system of units, as derived originally by Drake~\cite{Dra:71a} 
($ \mu_{B} = \frac{e \hbar}{2 mc} $ ).

In equation~\eref{eq:M1_operator_1}, one can recognize through the first two terms
$ (\bi {L} +2 \bi {S} = \bi {J} +  \bi {S})$ the usual form of the magnetic dipole transition 
operator~\cite{Cow:81a}. The other terms have been derived by Drake~\cite{Dra:71a} as the 
$O(\alpha^2 Z^2)$ corrections to the usual definition of the M1 operator in the Pauli approximation. 
Some confusion propagated in the literature due to embarassing misprints\footnote{
The two contributions in $\bi {r}_{i}\wedge \bi {p}_{j}$ and $\bi {r}_{j}\wedge \bi {p}_{i}$  
appearing with the same $(-)$
sign in the last term of~\eref{eq:M1_operator_1} do appear with opposite signs $(-)/(+)$ in 
Eissner and Zeippen~\cite{EisZei:81a} and Eissner~\cite{Eis:91a}.
Moreover the factor $(1/2)$  for the last two terms of \eref{eq:M1_operator_1}, is 
missing in ref.~\cite{Eis:91a}.
}
 but the original version~\cite{Dra:71a} should be the definitive one~\cite{Dra:07a}. \\

This operator, written in atomic units $(\mu_B = \alpha/2)$, splits into its one- and two-body 
components
\begin{equation}
\label{eq:M1_operator_split}
  \bi {Q}=\sum_{i=1}^{N}\bi {Q}_{i} + \sum_{i<j}^{N}\bi {Q}_{ij}\;,
\end{equation}
with
\begin{eqnarray}
\label{eq:M1_1}
\fl \bi{Q}_{i}= \mu_{B}\Bigg\{(\bi {l}_{i}+2\bi
{s}_{i})\left[1+\frac{\alpha^{2}}{2}\left(\frac{\partial^{2}}{\partial
r_{i}^{2}}+\frac{2}{r_{i}}\frac{\partial}{\partial r_{i}
}-\frac{\bi {l}_i^{2}}{r^{2}_{i}}-\frac{\epsilon^{2}}{80}
r_{i}^{2}\right)\right]\nonumber\\
\lo+ \frac{\alpha^{2}}{2}\left[\bi {p}_{i}\wedge (\bi
{p}_{i}\wedge \bi {s}_{i})+2\bi {r}_{i}\wedge (\bi {r}_{i}\wedge
\bi
{s}_{i})\left(\frac{\epsilon^{2}}{80}-\frac{Z}{r_{i}^{3}}\right)+\frac{\epsilon}{4}\bi
{s_{i}}\right]\Bigg\}
\end{eqnarray}
and
\begin{eqnarray}
\label{eq:M1_2}
\fl \bi {Q}_{ij}= \mu_{B}\alpha^{2}\Bigg[\frac{\bi {r}_{ij}\wedge
[\bi {r}_{ij}\wedge (\bi {s}_{i}+\bi
{s}_{j})]}{r_{ij}^{3}}+\frac{1}{2}\frac{(\bi {r}_{i}\wedge \bi
{r}_{j})\bi {r}_{ij}\cdot (\bi {p}_{i}+\bi {p}_{j})}{r_{ij}^{3}} -
\frac{1}{2}\frac{\bi {r}_{i}\wedge \bi {p}_{j}+\bi {r}_{j}\wedge
\bi {p}_{i}}{r_{ij}}\Bigg]
\end{eqnarray}
where $\epsilon \equiv \Delta E_{ul}$ is the transition energy in hartrees.

\section{Derivation of the irreducible tensorial form}

\subsection{The one-body operator}

The vectors $\bi {A}_{i}$ and $ \bi {B}_{i}$ defined as
\begin{equation}
\label{eq:A_def}
  \bi {A}_{i} \equiv \bi {p}_{i}\wedge (\bi {p}_{i}\wedge \bi {s}_{i})
  = \bi {p}_{i}(\bi {p}_{i}\cdot \bi {s}_{i})
  - \bi {s}_{i}(\bi {p}_{i}\cdot \bi {p}_{i}) \, ,
\end{equation}
and
\begin{equation}
\label{eq:B_def}
\bi {B}_{i} \equiv \bi {r}_{i}\wedge (\bi {r}_{i}\wedge \bi {s}_{i})
= \bi {r}_{i}(\bi {r}_{i}\cdot \bi {s}_{i})
  - \bi {s}_{i}(\bi {r}_{i}\cdot \bi {r}_{i}) \, ,
\end{equation}
appearing in the one-body operator \eref{eq:M1_1},
can be rewritten in irreducible tensor form 
\begin{equation}
\label{eq:A_tens_1}
\fl \bi {A}_{i}=\frac{2\sqrt{3}}{3} \left[\bi {s}_{i}^{(1)}\times
\left(\bi {p}_{i}^{(1)}\times \bi
{p}_{i}^{(1)}\right)^{(0)}\right]^{(1)}-\frac{\sqrt{15}}{3}
\left[\bi {s}_{i}^{(1)}\times \left(\bi {p}_{i}^{(1)}\times \bi
{p}_{i}^{(1)}\right)^{(2)}\right]^{(1)} \; , 
\end{equation}

\begin{equation}
\label{eq:B_tens_1}
\fl\bi {B}_{i}=\frac{2\sqrt{3}}{3} \left[\bi {s}_{i}^{(1)}\times
\left(\bi {r}_{i}^{(1)}\times \bi
{r}_{i}^{(1)}\right)^{(0)}\right]^{(1)}-\frac{\sqrt{15}}{3}
\left[\bi {s}_{i}^{(1)}\times \left(\bi {r}_{i}^{(1)}\times \bi
{r}_{i}^{(1)}\right)^{(2)}\right]^{(1)} \; , 
\end{equation}
by applying  the angular momentum theory and operator techniques 
\cite{BriSat:68a,Jud:98a,Kar:96a} and using the recoupling formula 
for commuting irreducible tensors \cite{Varetal:88a}.
%\subsubsection{Tensorial form of $\bi {A}_{i}$ } \ \\
From the  tensorial form of the linear momentum 
%$\bi {p} = -i \hbar \bi {\bnabla}$ 
  $\bi {p} = -i \bi {\bnabla}$ (in a.u.)~\cite{Jud:98a}

\begin{equation}
\label{eq:linear_momentum_tensor}
\bi {p}^{(1)} = -\frac{i}{r}
\bnabla_{\Omega}^{(1)} - i \bi C^{(1)}
\frac{\partial}{\partial r} \; ,
\end{equation}
the tensorial product $\left(\bi {p}_{i}^{(1)}\times \bi
{p}_{i}^{(1)}\right)^{(2)}$ is written in terms of the angular part $\bnabla_{\Omega}^{(1)}$ 
of the differential operator $\bnabla$ and the renormalized  spherical harmonic 
operator~$\bi C^{(1)}$:

\begin{eqnarray}
\label{eq:p_cross_p_k=2}
 \fl \left(\bi {p}_{i}^{(1)}\times \bi
{p}_{i}^{(1)}\right)^{(2)}&=&-\frac{1}{r_{i}^{2}}
\left(\bnabla_{\Omega_{i}}^{(1)}\times \bnabla_{\Omega_{i}}^{(1)}
\right)^{(2)}-\frac{1}{r_{i}} \frac{\partial}{\partial r_{i} }
\left(\bnabla_{\Omega_{i}}^{(1)}\times \bi {C}_{i}^{(1)}
\right)^{(2)}\nonumber\\
\nonumber\\
& &- \frac{\partial}{\partial r_{i} } \frac{1}{r_{i}} \left( \bi
{C}_{i}^{(1)}\times \bnabla_{\Omega_{i}}^{(1)}
\right)^{(2)}-\frac{\partial^{2}}{\partial r_{i}^{2}} \left( \bi
{C}_{i}^{(1)}\times \bi {C}_{i}^{(1)} \right)^{(2)} \, .
\end{eqnarray}
Defining the commutator of an irreducible tensor product as in~\cite{Varetal:88a}
\begin{equation}
\label{eq:Commutator_tenros_product_def}
\fl \mathfrak{R}^{k_{1}k_{2}}_{k q} ( \bi {A}^{(k_{1})} , \bi {B}^{(k_{2})})
=
\left(\bi {A}^{(k_{1})} \times \bi {B}^{(k_{2})}\right)^{(k)}_{ \; q}
 - (-1)^{k_{1}+k_{2}-k}
 \left(\bi {B}^{(k_{2})} \times \bi {A}^{(k_{1})} \right)^{(k)}_{ \; q} \, ,
\end{equation}
the two tensors $\left(\nabla_{\Omega_{i}}^{(1)}\times \bi {C}_{i}^{(1)} \right)^{(2)}$
and $\left( \bi {C}_{i}^{(1)}\times \nabla_{\Omega_{i}}^{(1)} \right)^{(2)}$
appearing in the second and third terms of \eref{eq:p_cross_p_k=2} 
are related through their commutator\footnote{\label{Rk1k2} 
We can see the general expression of the 
$\mathfrak{R}^{1k_{2}}_{k} ( \bnabla_{\Omega}^{(1)} , \bi {C}^{(k_{2})})$ 
in the book of Varshalovich and al \cite{Varetal:88a} (equation 113 page 497).} 
that reduces to 
\begin{equation}
\label{eq:commutator_nabla_C1}
\fl \mathfrak{R}^{11}_{2}(\bnabla_{\Omega}^{(1)} , \bi
{C}^{(1)})=\left(\bnabla_{\Omega}^{(1)}\times \bi
{C}^{(1)}\right)^{(2)}-(-1)^{1+1-2}\left(\bi {C}^{(1)}\times
\bnabla_{\Omega}^{(1)} \right)^{(2)} = - \frac{\sqrt{6}}{3} \bi
{C}^{(2)} \, .
\end{equation}
The fourth term of \eref{eq:p_cross_p_k=2} is simplified by using the key reduction 
formula  \eref{CK} for the tensor product of spherical harmonics \cite{Jud:98a}:
\begin{equation}
\label{CK}
\left(\bi {C}^{(k_{1})} \times \bi
{C}^{(k_{2})}\right)^{(k)}=(-1)^{(k)} \sqrt{2k+1} \left(
\begin{array}{ccc}
k_{1} & k & k_{2} \\
0 & 0 & 0
\end{array}
\right) \bi {C}^{(k)} \; .
\end{equation}
Using equations \eref{eq:commutator_nabla_C1} and \eref{CK}, expression~\eref{eq:p_cross_p_k=2} 
becomes
\begin{eqnarray}
\label{eq:p_cross_p_k=2_B}
\fl \left(\bi {p}_{i}^{(1)}\times \bi
{p}_{i}^{(1)}\right)^{(2)}=&-\frac{1}{r_{i}^{2}}
\left(\bnabla_{\Omega_{i}}^{(1)}\times \bnabla_{\Omega_{i}}^{(1)}
\right)^{(2)}-\left(\frac{1}{r_{i}} \frac{\partial}{\partial
r_{i}}+\frac{\partial}{\partial r_{i}} \frac{1}{r_{i}}\right)
 \left( \bi {C}_{i}^{(1)}\times \bnabla_{\Omega_{i}}^{(1)}
\right)^{(2)}\nonumber\\
& +\frac{\sqrt{6}}{3} \left(\frac{1}{r_{i}}
\frac{\partial}{\partial r_{i}}-\frac{\partial^{2}}{\partial
r_{i}^{2}}\right) \bi {C}_{i}^{(2)} \, .
\end{eqnarray}
The tensor product $\left( \bi {p}_{i}^{(1)}\times \bi {p}_{i}^{(1)} \right)^{(0)}$ appearing 
in \eref{eq:A_tens_1} is simply obtained thanks to the relation with the scalar product 
$\bi {p}_{i} \cdot \bi {p}_{i}$ :
\begin{equation}
\label{eq:p_cross_p_k=0}
\fl
\left(\bi {p}_{i}^{(1)}\times \bi {p}_{i}^{(1)}\right)^{(0)}
= - \frac{1}{\sqrt{3}} \bi {p}_{i} \cdot \bi {p}_{i} = \frac{1}{\sqrt{3}} \triangle_{i}
= + \frac{\sqrt{3}}{3}
\left(\frac{\partial^{2}}{\partial r_{i}^{2}} + \frac{2}{r_{i}}
\frac{\partial}{\partial r_{i} } - \frac{\bi l_{i}^{2}}{r_{i}^{2}}
\right)
\end{equation}
By introducing equations \eref{eq:p_cross_p_k=0} and \eref{eq:p_cross_p_k=2_B} 
in \eref{eq:A_tens_1}, one gets the final tensorial expression for~$ \bi {A}_{i} $:
\begin{eqnarray}
\label{eq:A_tens_2}
 \bi {A}_{i} =
 & +  \frac{2}{3} \left(\frac{\partial^{2}}{\partial
r_{i}^{2}} + \frac{2}{r_{i}} \frac{\partial}{\partial r_{i} } -
\frac{\bi l_{i}^{2}}{r_{i}^{2}} \right) \bi {s}_{i}^{(1)}-
\frac{\sqrt{10}}{3} \left(\frac{1}{r_{i}} \frac{\partial}{\partial
r_{i}}-\frac{\partial^{2}}{\partial r_{i}^{2}}\right)\left(\ \bi
{s}_{i}^{(1)}\times \bi {C}_{i}^{(2)}\right)^{(1)}\nonumber\\
& + \frac{\sqrt{15}}{3} \frac{1}{r_{i}^{2}} \left[\ \bi
{s}_{i}^{(1)}\times \left(\bnabla_{\Omega_i}^{(1)} \times
\bnabla_{\Omega_i}^{(1)}\right)^{(2)}\right]^{(1)}\nonumber\\
& + \frac{\sqrt{15}}{3} \left(\frac{1}{r_{i}}
\frac{\partial}{\partial r_{i}}+\frac{\partial}{\partial r_{i}}
\frac{1}{r_{i}}\right) \left[\bi {s}_{i}^{(1)}\times \left(\bi
{C}_{i}^{(1)}\times \bnabla_{\Omega_{i}}^{(1)}
\right)^{(2)}\right]^{(1)}   \, .
\end{eqnarray}
The tensorial form \eref{eq:B_tens_1} of  $ \bi {B}_{i} $ reduces to 
\begin{equation}
\label{eq:B_tens_2}
\bi {B}_{i}=-\frac{2}{3} r_{i}^{2} \bi
{s}_{i}^{(1)}-\frac{\sqrt{10}}{3} \; r_{i}^{2} \left(\bi
{s}_{i}^{(1)}\times \bi {C}_{i}^{(2)}\right)^{(1)} \; ,
\end{equation}
 by using $ \bi r ^{(1)} = r \, \bi {C}^{(1)} $ and \eref{CK}. \\
 
 \noindent Inserting expressions \eref{eq:A_tens_2} and \eref{eq:B_tens_2} into 
 equation \eref{eq:M1_1}, one obtains, after some regrouping, the tensorial form
of the one-body $M1$ transition operator:
\begin{eqnarray}
\label{eq:M1_1_tens}
\sum_{i=1}^{N} \bi {Q}_{i}= &\mu_{B}  \sum_{i=1}^{N}
\Bigg\{\left[1+\frac{\alpha^{2}}{2}\left(\frac{\partial^{2}}{\partial
r_{i}^{2}}+\frac{2}{r_{i}}\frac{\partial}{\partial r_{i}
}-\frac{\bi l_{i}^{2}}{r^{2}_{i}}-\frac{\epsilon^{2}}{80}
r_{i}^{2}\right)\right] \bi {l}_{i}^{(1)}\nonumber\\
&+\left[2+ \frac{4}{3}
\alpha^{2}\left(\frac{\partial^{2}}{\partial
r_{i}^{2}}+\frac{2}{r_{i}}\frac{\partial}{\partial r_{i}
}-\frac{l_{i}^{2}}{r^{2}_{i}}-\frac{\epsilon^{2}}{64}
r_{i}^{2}+\frac{Z}{2r_{i}}+\frac{3}{32} \epsilon \right)\right]\bi {s}_{i}^{(1)}\nonumber\\
&+\frac{\sqrt{10}}{6} \alpha^{2}\left[\frac{\partial^{2}}{\partial
r_{i}^{2}}-\frac{1}{r_{i}}\frac{\partial}{\partial
r_{i}}-r_{i}^{2}\frac{\epsilon^{2}}{40}
+ \frac{2Z}{r_{i}}\right]\left[\bi
{s}_{i}^{(1)}\times \bi {C}_{i}^{(2)}\right]^{(1)}\nonumber\\
&+ \frac{\sqrt{15}}{6} \alpha^{2} \frac{1}{r_{i}^{2}}\left[\ \bi
{s}_{i}^{(1)}\times \left(\bnabla_{\Omega_i}^{(1)} \times
\bnabla_{\Omega_i}^{(1)}\right)^{(2)}\right]^{(1)}\nonumber\\
&+ \frac{\sqrt{15}}{6} \alpha^{2}\left[\frac{1}{r_{i}}
\frac{\partial}{\partial r_{i}}+\frac{\partial}{\partial r_{i}}
\frac{1}{r_{i}}\right] \left[\bi {s}_{i}^{(1)}\times \left(\bi
{C}_{i}^{(1)}\times \bnabla_{\Omega_{i}}^{(1)}
\right)^{(2)}\right]^{(1)}\Bigg\} \; .
\end{eqnarray}

\subsection{The two-body operator}

For the two-body magnetic dipole transition operator~\eref{eq:M1_2}, we need
to investigate the irreducible tensorial expressions of the three following contributions:
\begin{eqnarray}
\label{CDE}
\bi {C_{ij}}&=& \frac{\bi {r}_{ij}\wedge \bi {r}_{ij}\wedge (\bi
{s}_{i}+\bi {s}_{j})}{r_{ij}^{3}} \; , \nonumber\\
\bi {D_{ij}}&=& \frac{(\bi {r}_{i}\wedge \bi {r}_{j}) \bi
{r}_{ij}\cdot (\bi {p}_{i}+\bi {p}_{j})}{r_{ij}^{3}} \; , \\
\bi {E_{ij}}&=&\frac{\bi {r}_{i}\wedge \bi {p}_{j}+\bi
{r}_{j}\wedge \bi {p}_{i}}{r_{ij}} \; .  \nonumber
\end{eqnarray}

\subsubsection{Tensorial form of  $\bi {C}_{ij}$} \ \\

Similarly to equations \eref{eq:A_tens_1} and \eref{eq:B_tens_1}, the tensorial expression 
of the double vector product appearing in $\bi {C_{ij}}$ becomes
\begin{eqnarray}
\label{C_ij tens1}
\bi {C_{ij}}=&+\frac{2\sqrt{3}}{3} \frac{1}{r_{ij}^{3}}
\left[(\bi {s}_{i}^{(1)}+\bi {s}_{j}^{(1)})\times \left(\bi
{r}_{ij}^{(1)}\times \bi
{r}_{ij}^{(1)}\right)^{(0)}\right]^{(1)}\nonumber\\
&-\frac{\sqrt{15}}{3} \frac{1}{r_{ij}^{3}}
\left[
(\bi {s}_{i}^{(1)} + \bi {s}_{j}^{(1)}) \times
  \left(\bi {r}_{ij}^{(1)} \times \bi {r}_{ij}^{(1)} \right)^{(2)}
  \right]^{(1)} \, .
\end{eqnarray}

\noindent
Let be $\bi {X}_{ij}$ and $\bi {Y}_{ij}$, the first and the second terms of~\eref{C_ij tens1} 
ie., $ \bi {C_{ij}} = \bi {X}_{ij} + \bi {Y}_{ij}$.

\paragraph{Calculation of $ \bi {X}_{ij} $} \ \\

Combining the vectorial form of the tensor product of rank zero 
\begin{equation}
\label{r1xr1_zero}
\left(\bi {r}_{ij}^{(1)}\times \bi
{r}_{ij}^{(1)}\right)^{(0)}= - \frac{1}{\sqrt{3}} \; \bi {r}_{ij} \cdot
\bi {r}_{ij} = -\frac{1}{\sqrt{3}} \; r_{ij}^{2}
\end{equation}
with the well-known expression \cite{Jud:98a}
\begin{equation}
\label{one_over_rij}
 \frac{1}{r_{ij}}= \sum_{k} (-1)^{k}
\frac{r_{<}^{k}}{r_{>}^{k+1}} (2k+1)^{1/2} \left(\bi
{C}_{i}^{(k)}\times \bi {C}_{j}^{(k)}\right)^{(0)}
\, ,
\end{equation}
the sum over the electron pairs of the ${X}_{ij}$ contributions becomes
\begin{equation}
\label{Xij_tens_3}
\fl\sum_{i<j}\bi {X}_{ij}=-\frac{2}{3} \sum_{i\neq j} \sum_{k}
(-1)^{k} \sqrt{(2k+1)} \frac{r_{j}^{k}}{r_{i}^{k+1}}
\varepsilon(r_{i}-r_{j})
 \left[(\bi {s}_{j}^{(1)}+\bi {s}_{i}^{(1)})
\times \left(\bi {C}_{j}^{(k)}\times \bi
{C}_{i}^{(k)}\right)^{(0)}\right]^{(1)}
\end{equation}
with 
\begin{equation}
\label{epsilon_def}
 \varepsilon(r_{j}-r_{i})=\left\{
\begin{array}{ccc}
1 & \mbox{si}& r_{i}<r_{j}\\
0 & \mbox{si}& r_{i}>r_{j} \; .
\end{array}
\right.
\end{equation}
%done

\paragraph{Calculation of $ \bi {Y}_{ij} $} \ \\

Starting from $ \bi r ^{(1)} = r \, \bi {C}^{(1)} $ and $ \bi r_{ij} \equiv \bi r_i - \bi r_j $, 
one first builds the irreducible tensor of rank two
\begin{equation}
\label{eq:r_ij_def}
\left(\bi {r}_{ij}^{(1)}\times \bi
{r}_{ij}^{(1)}\right)^{(2)}=\sqrt{\frac{2}{3}} \left(r_{i}^{2}\bi
{C}_{i}^{(2)}+r_{j}^{2}\bi {C}_{j}^{(2)}\right)-2 \; r_{i}r_{j}
\left(\bi {C}_{i}^{(1)}\times \bi {C}_{j}^{(1)}\right)^{(2)}
\end{equation}
using \eref{CK}. Taking the tensorial form of $r_{ij}^{-3}$ \cite{Jud:98a}
\begin{equation}
\label{eq:r_ij_minus3}
\frac{1}{r_{ij}^{3}}=\frac{1}{r_{>}^{2}-r_{<}^{2}} \sum_{k}
(-1)^{k} \frac{r_{<}^{k}}{r_{>}^{k+1}} (2k+1)^{3/2} \left(\bi
{C}_{i}^{(k)}\times \bi {C}_{j}^{(k)}\right)^{(0)}
\, ,
\end{equation}
the $\bi {Y}_{ij} $ contribution  becomes
\begin{eqnarray}
\label{Y_ij_tens_1}
\fl
 \bi {Y}_{ij}= -\frac{\sqrt{10}}{3}
\frac{r_{i}^{2}}{r_{>}^{2}-r_{<}^{2}} \sum_{k} (-1)^{k}
\frac{r_{<}^{k}}{r_{>}^{k+1}} (2k+1)^{3/2} \nonumber\\
\times \Bigg\{\left[(\bi {s}_{i}^{(1)}+\bi {s}_{j}^{(1)})\times
\bi {C}_{i}^{(2)}\right]^{(1)}\times \left[\bi {C}_{i}^{(k)}\times
\bi
{C}_{j}^{(k)}\right]^{(0)}\Bigg\}^{(1)}\nonumber\\
\fl \quad \quad \quad -\frac{\sqrt{10}}{3} \frac{r_{j}^{2}}{r_{>}^{2}-r_{<}^{2}}
\sum_{k} (-1)^{k} \frac{r_{<}^{k}}{r_{>}^{k+1}} (2k+1)^{3/2}\nonumber\\
\times \Bigg\{\left[(\bi {s}_{j}^{(1)}+\bi {s}_{i}^{(1)})\times
\bi {C}_{j}^{(2)}\right]^{(1)}\times \left[\bi {C}_{j}^{(k)}\times
\bi
{C}_{i}^{(k)}\right]^{(0)}\Bigg\}^{(1)}\nonumber\\
\fl \quad \quad \quad +\frac{2}{3}\sqrt{15} \frac{r_{i}r_{j}}{r_{>}^{2}-r_{<}^{2}}
\sum_{k} (-1)^{k} \frac{r_{<}^{k}}{r_{>}^{k+1}} (2k+1)^{3/2}\nonumber\\
 \times \Bigg\{\left[(\bi {s}_{i}^{(1)}+\bi {s}_{j}^{(1)})\times
\left(\bi {C}_{i}^{(1)}\times \bi
{C}_{j}^{(1)}\right)^{(2)}\right]^{(1)}\times \left[\bi
{C}_{i}^{(k)}\times \bi {C}_{j}^{(k)}\right]^{(0)}\Bigg\}^{(1)}
\, .
\end{eqnarray}
The calculation of the  three terms is rather tedious, requiring numerous recouplings for 
regrouping the spin and angular factors. Expressions of these terms are given  
in \ref{app2}. The thereby defined and calculated $\bi {F}_{ij}$ and $\bi {H}_{ij}$ 
contributions are relevant to the first and third term of \eref{Y_ij_tens_1}, respectively, 
while the second term is obtained from the first one by merely interchanging the $(ij)$ indices. 
After some further effort in regrouping similar contributions, the sum over all electron 
pairs of \eref{Y_ij_tens_1} reduces to
\begin{eqnarray}
\label{Y_ij_tens_2}
\fl  \sum_{i<j}\bi {Y}_{ij}=
- \frac{1}{3} \sum_{i\neq j} \sum_{k}
(-1)^{k} \sqrt{\frac{(2k+1)(k+1)}{2k+3}}
\frac{r_{j}^{k}}{r_{i}^{k+1}} \varepsilon(r_{i}-r_{j})\nonumber\\
\fl \quad \quad \quad \times \Bigg\{2\sqrt{\frac{2k}{2k-1}} \left[(\bi
{s}_{j}^{(1)}+\bi {s}_{i}^{(1)})\times \left(\bi
{C}_{j}^{(k)}\times \bi {C}_{i}^{(k)}\right)^{(2)}\right]^{(1)}\nonumber\\
\fl \quad \quad \quad \quad
+ \left(1-\frac{r_{j}^{2}}{r_{i}^{2}}\right) \sqrt{3(2k+5)(k+2)}
\left[(\bi {s}_{j}^{(1)}+\bi {s}_{i}^{(1)})\times \left(\bi
{C}_{j}^{(k)}\times \bi
{C}_{i}^{(k+2)}\right)^{(2)}\right]^{(1)}\Bigg\}
\, .
\nonumber\\
\end{eqnarray}
The final two-body contribution $\sum_{i<j}\bi {C}_{ij}$, as defined in  \eref{CDE}, are 
calculated according to equations  \eref{Xij_tens_3} and \eref{Y_ij_tens_2}:
\begin{eqnarray}
\label{eq:sum_ij_C_ij_final}
\fl \quad  \quad \sum_{i<j}\bi {C}_{ij}= - \frac{2}{3} \sum_{i\neq j} \sum_{k}
(-1)^{k} \sqrt{(2k+1)} \frac{r_{j}^{k}}{r_{i}^{k+1}}
\varepsilon(r_{i}-r_{j})\nonumber\\
\quad \quad \quad  \times \left[(\bi {s}_{j}^{(1)}+\bi {s}_{i}^{(1)})\times
\left(\bi
{C}_{j}^{(k)}\times \bi {C}_{i}^{(k)}\right)^{(0)}\right]^{(1)}\nonumber\\
- \frac{2\sqrt{2}}{3} \sum_{i\neq j} \sum_{k} (-1)^{k}
\sqrt{\frac{k(2k+1)(k+1)}{(2k+3)(2k-1)}}
\frac{r_{j}^{k}}{r_{i}^{k+1}} \varepsilon(r_{i}-r_{j})\nonumber\\
\quad \quad \quad \times \left[(\bi {s}_{j}^{(1)}+\bi {s}_{i}^{(1)})\times
\left(\bi
{C}_{j}^{(k)}\times \bi {C}_{i}^{(k)}\right)^{(2)}\right]^{(1)}\nonumber\\
- \frac{\sqrt{3}}{3} \sum_{i\neq j} \sum_{k} (-1)^{k}
\sqrt{\frac{(2k+5)(2k+1)(k+2)(k+1)}{2k+3}}
\frac{r_{j}^{k}}{r_{i}^{k+1}}
\varepsilon(r_{i}-r_{j})\nonumber\\
\quad \quad \quad \times \left(1-\frac{r_{j}^{2}}{r_{i}^{2}}\right) \left[(\bi
{s}_{j}^{(1)}+\bi {s}_{i}^{(1)})\times \left(\bi
{C}_{j}^{(k)}\times \bi
{C}_{i}^{(k+2)}\right)^{(2)}\right]^{(1)} \; .
\end{eqnarray}

  \subsubsection{Tensorial form of  $\bi {D}_{ij}$} \ \\

    There are different ways for getting the tensorial expression of the two-body 
    contribution $\bi {D}_{ij}$ appearing in \eref{CDE}. One approach is to evaluate 
    the two subparts  $ \bi {r}_{ij}\cdot (\bi {p}_{i}+\bi {p}_{j}) $ and 
    $r_{ij}^{-3} ( \bi {r}_{i}\wedge \bi {r}_{j} ) $ of the operator  separately. 
    Using equations \eref{eq:linear_momentum_tensor}, \eref{CK} and the expression 
    of  $\mathfrak{R}^{1k}_{k-1}(\bnabla_{\Omega}^{(1)} ,
\bi {C}^{(k)})$ (see the footnote \ref{Rk1k2}) , one easily obtains
\begin{eqnarray}
\label{r_ij_cdot_pi+pj}
\fl \bi {r}_{ij}\cdot (\bi {p}_{i}+\bi {p}_{j})=&i\left(r_{j}
\frac{\partial}{\partial r_{j}}-r_{i} \frac{\partial}{\partial
r_{i}}\right)+i\sqrt{3} \left(r_{i} \frac{\partial}{\partial
r_{j}}-r_{j} \frac{\partial}{\partial r_{i}}\right) \left(\bi
{C}_{i}^{(1)}\times \bi {C}_{j}^{(1)}\right)^{(0)}\nonumber\\
&+i\sqrt{3}\frac{r_{i}}{r_{j}} \left( \bi {C}_{i}^{(1)} \times
\bnabla_{\Omega_{j}}^{(1)}
\right)^{(0)}-i\sqrt{3}\frac{r_{j}}{r_{i}} \left( \bi
{C}_{j}^{(1)} \times \bnabla_{\Omega_{i}}^{(1)} \right)^{(0)}
\, .
\end{eqnarray}
From the rank one tensor associated to the vector product 
\begin{equation}
\label{r_cross_r}
 \bi {r}_{i}\wedge \bi {r}_{j}= - i \sqrt{2} \; r_{i}r_{j} \left(\bi
{C}_{i}^{(1)}\times \bi {C}_{j}^{(1)}\right)^{(1)}\;
\end{equation}
and from \eref{eq:r_ij_minus3} and \eref{CCCC(kk)0(11)1}, we get
\begin{equation}
\label{r_i_cross_rj_over_rij_m3_2}
 \frac{\bi {r}_{i}\wedge \bi
{r}_{j}}{r_{ij}^{3}}=i\frac{\sqrt{3}}{3} \sum_{k} (-1)^{k}
\frac{r_{<}^{k}}{r_{>}^{k+1}} \sqrt{(2k+1)(k+1)k} \left(\bi
{C}_{i}^{(k)}\times \bi {C}_{j}^{(k)}\right)^{(1)}
\, .
\end{equation}
By recombining equations \eref{r_i_cross_rj_over_rij_m3_2} and \eref{r_ij_cdot_pi+pj} 
for building the $\bi D_{ij}$ operator, one gets
\begin{eqnarray}
\label{D_ij_tens_1}
\fl \bi {D_{ij}}
= \frac{(\bi {r}_{i}\wedge \bi
{r}_{j})}{r_{ij}^{3}}
 \bi {r}_{ij}\cdot (\bi {p}_{i}+\bi
{p}_{j})\nonumber\\
\fl \qquad
=
-\frac{\sqrt{3}}{3} \sum_{k} (-1)^{k}
\frac{r_{<}^{k}}{r_{>}^{k+1}} \left(r_{j} \frac{\partial}{\partial
r_{j}}-r_{i} \frac{\partial}{\partial r_{i}}\right)
\sqrt{(2k+1)(k+1)k} \;  \left(\bi
{C}_{i}^{(k)}\times \bi {C}_{j}^{(k)}\right)^{(1)}   \nonumber\\
\fl\qquad\quad-\sum_{k} (-1)^{k} \frac{r_{<}^{k}}{r_{>}^{k+1}}
\left(r_{i} \frac{\partial}{\partial r_{j}}-r_{j}
\frac{\partial}{\partial
r_{i}}\right) \sqrt{(2k+1)(k+1)k} \nonumber\\
\qquad\qquad\qquad\qquad\times \left[\left(\bi
{C}_{i}^{(k)}\times \bi {C}_{j}^{(k)}\right)^{(1)}\times \left(\bi
{C}_{i}^{(1)}\times
\bi {C}_{j}^{(1)}\right)^{(0)}\right]^{(1)}   \nonumber\\
\fl\qquad\quad-\sum_{k} (-1)^{k} \frac{r_{i}}{r_{j}}
\frac{r_{<}^{k}}{r_{>}^{k+1}} \sqrt{(2k+1)(k+1)k}  \nonumber\\
 \qquad \qquad \qquad \qquad \times\left[\left(\bi
{C}_{i}^{(k)}\times \bi {C}_{j}^{(k)}\right)^{(1)}\times \left(\bi
{C}_{i}^{(1)}\times
\bnabla_{\Omega_{j}}^{(1)}\right)^{(0)}\right]^{(1)}\nonumber\\
\fl\qquad\quad+\sum_{k} (-1)^{k} \frac{r_{j}}{r_{i}}
\frac{r_{<}^{k}}{r_{>}^{k+1}} \sqrt{(2k+1)(k+1)k}\nonumber\\
 \qquad \qquad \qquad \qquad \times\left[\left(\bi
{C}_{i}^{(k)}\times \bi {C}_{j}^{(k)}\right)^{(1)}\times \left(\bi
{C}_{j}^{(1)}\times
\bnabla_{\Omega_{i}}^{(1)}\right)^{(0)}\right]^{(1)}
\, .
\end{eqnarray}
Let us denote the above four contributions to $ \bi D_{ij}$ as 
$ \bi K_{ij}$, $ \bi L_{ij}$, $ \bi M_{ij}$ and $ \bi N_{ij}$,
respectively. The tensor products of the second and third terms are 
recoupled in equations \eref{CCCC(kk)1(11)0}
and \eref{CCCnabla(kk)1(11)0}. After some tedious work, the two first 
contributions to $\bi D_{ij}$ reduce to
\begin{eqnarray}
\label{K_ij_plus_L_ij_tens_1}
\fl\bi {K}_{ij}+\bi {L}_{ij}=&&\frac{\sqrt{3}}{3} \sum_{k}
(-1)^{k} \frac{r_{<}^{k}}{r_{>}^{k+1}} \sqrt{(2k+1)(k+1)k}\nonumber\\
&& \times \Bigg[\left(r_{i} \frac{r_{<}}{r_{>}}
\frac{k+2}{2k+3}+r_{i} \frac{r_{>}}{r_{<}}
\frac{k-1}{2k-1}-r_{j}\right) \frac{\partial}{\partial r_{j}}
\left(\bi
{C}_{i}^{(k)}\times \bi {C}_{j}^{(k)}\right)^{(1)}\nonumber\\
\nonumber\\
&&+\left(r_{j} \frac{r_{<}}{r_{>}} \frac{k+2}{2k+3}+r_{j}
\frac{r_{>}}{r_{<}} \frac{k-1}{2k-1}-r_{i}\right)
\frac{\partial}{\partial r_{i}} \left(\bi {C}_{j}^{(k)}\times \bi
{C}_{i}^{(k)}\right)^{(1)}\Bigg] \, . \nonumber\\
\end{eqnarray}
The sum of the last two terms of \eref{K_ij_plus_L_ij_tens_1} being symmetric 
with respect to $(ij$)-exchange, the
summation over all pairs becomes
\begin{eqnarray}
\label{K_ij_plus_L_ij_tens_2}
\fl\sum_{i<j} (\bi {K}_{ij}+\bi {L}_{ij})=&\frac{\sqrt{3}}{3}
\sum_{i\neq j} \sum_{k}
(-1)^{k} \frac{r_{<}^{k}}{r_{>}^{k+1}} \sqrt{(2k+1)(k+1)k}\nonumber\\
&\times  \left(r_{<}^{2} \frac{k+2}{2k+3}+r_{>}^{2}
 \frac{k-1}{2k-1}-r_{j}^{2}\right)
\frac{1}{r_{j}} \frac{\partial}{\partial r_{i}} \left(\bi
{C}_{i}^{(k)}\times \bi {C}_{j}^{(k)}\right)^{(1)}
\, .
\end{eqnarray}
Remembering that
$\left(\bi {C}_{i}^{(k)}\times \bi {C}_{j}^{(k)}\right)^{(1)} = 
- \left(\bi {C}_{j}^{(k)}\times \bi
{C}_{i}^{(k)}\right)^{(1)}$, one realizes that the third and fourth terms 
of \eref{D_ij_tens_1}
are also symmetric with respect to $(ij$)-exchange, ie. $\bi {M}_{ij} = \bi {N}_{j \, i}$, 
leading to
\[
\sum_{i<j} \;(\bi {M}_{ij} + \bi {N}_{ij} ) =
\sum_{i \neq j} \; \bi {M}_{ij} =
\sum_{i \neq j} \; \bi {N}_{ij}  \, .
\]
The final expression for $\sum_{i<j} \, \bi {D}_{ij}$ is
\begin{eqnarray}
\label{D_ij_tens_2}
\fl\;\sum_{i<j} \bi {D}_{ij}=\frac{\sqrt{3}}{3} \sum_{i\neq j}
\sum_{k} (-1)^{k} \frac{r_{<}^{k}}{r_{>}^{k+1}}
\sqrt{(2k+1)(k+1)k}
  \left(r_{<}^{2}
\frac{k+2}{2k+3}+r_{>}^{2}
 \frac{k-1}{2k-1}-r_{j}^{2}\right)
\frac{1}{r_{j}} \frac{\partial}{\partial r_{j}}\nonumber\\
\qquad\qquad\qquad\qquad\qquad\qquad\qquad\qquad\times\left(\bi
{C}_{i}^{(k)}\times \bi {C}_{j}^{(k)}\right)^{(1)}\nonumber\\
\fl\qquad\qquad-\frac{\sqrt{3}}{3} \sum_{i\neq j} \sum_{k}
(-1)^{k} \sqrt{\frac{(2k+1)(k+2)^{2}k}{2k+3}} \frac{r_{i}}{r_{j}}
\frac{r_{<}^{k+1}}{r_{>}^{k+2}}\nonumber\\
\qquad\qquad\qquad\qquad\qquad\times\left(\bi {C}_{i}^{(k)}\times
\left(\bi {C}_{j}^{(k+1)}\times
\bnabla_{\Omega_{j}}^{(1)}\right)^{(k)}\right)^{(1)}\nonumber\\
\fl\qquad\qquad+\frac{\sqrt{3}}{3} \sum_{i\neq j} \sum_{k}
(-1)^{k} \sqrt{\frac{(2k+1)(k+2)}{2k+3}} \frac{r_{i}}{r_{j}}
\frac{r_{<}^{k+1}}{r_{>}^{k+2}}\nonumber\\
\qquad\qquad\qquad\qquad\quad\times\left(\bi {C}_{i}^{(k)}\times
\left(\bi {C}_{j}^{(k+1)}\times
\bnabla_{\Omega_{j}}^{(1)}\right)^{(k+1)}\right)^{(1)}\nonumber\\
\fl\qquad\qquad-\frac{\sqrt{3}}{3} \sum_{i\neq j} \sum_{k}
(-1)^{k} \sqrt{\frac{k(2k+3)}{2k+1}} \frac{r_{i}}{r_{j}}
\frac{r_{<}^{k}}{r_{>}^{k+1}}\nonumber\\
\qquad\qquad\qquad\qquad\quad\quad\times\left(\bi
{C}_{i}^{(k+1)}\times \left(\bi {C}_{j}^{(k)}\times
\bnabla_{\Omega_{j}}^{(1)}\right)^{(k)}\right)^{(1)}\nonumber\\
\fl\qquad\qquad-\frac{\sqrt{3}}{3} \sum_{i\neq j} \sum_{k}
(-1)^{k} \sqrt{\frac{k^{2}(k+2)(2k+3)}{2k+1}} \frac{r_{i}}{r_{j}}
\frac{r_{<}^{k}}{r_{>}^{k+1}}\nonumber\\
\qquad\qquad\qquad\qquad\quad\times\left(\bi {C}_{i}^{(k+1)}\times
\left(\bi {C}_{j}^{(k)}\times
\bnabla_{\Omega_{j}}^{(1)}\right)^{(k+1)}\right)^{(1)} \, .
\end{eqnarray}

  \subsubsection{Tensorial form of  $\bi {E}_{ij}$} \ \\

The two terms appearing in $\bi E_{ij}$ \eref{CDE} being symmetric with respect 
to exchange $(ij)$,
one will restrict to one of the two. Using \eref{one_over_rij} and the tensorial 
form of the vector product, one easily obtains
\begin{eqnarray}
\label{ri_cross_pi_over_rij}
\fl \frac{\bi {r}_{i} \wedge \bi {p}_{j}}{r_{ij}}=-\sqrt{2}
\sum_{k} (-1)^{k} \sqrt{2k+1}
\frac{r_{<}^{k}}{r_{>}^{k+1}}\frac{r_{i}}{r_{j}}
\times\left(\left(\bi {C}_{i}^{(k)}\times \bi
{C}_{j}^{(k)}\right)^{(0)}\times \left(\bi {C}_{i}^{(1)}\times
\bnabla_{\Omega_{j}}^{(1)}\right)^{(0)}\right)^{(1)}\nonumber\\
\!\!\!\!\!\!\!\!\!\!-\sqrt{2} \sum_{k} (-1)^{k} \sqrt{2k+1}
\frac{r_{<}^{k}}{r_{>}^{k+1}} r_{i}\frac{\partial}{\partial r_{j}}
\times\left(\left(\bi {C}_{i}^{(k)}\times \bi
{C}_{j}^{(k)}\right)^{(0)}\times \left(\bi {C}_{i}^{(1)}\times \bi
{C}_{j}^{(1)}\right)^{(1)}\right)^{(1)} \, . \nonumber\\
\end{eqnarray}
The two tensor products are recoupled using equations \eref{CCCnabla(kk)1(11)0} 
and \eref{CCCC(kk)0(11)1} for getting the following form
\begin{eqnarray}
\label{RPR}
\fl\frac{\bi {r}_{i} \wedge \bi {p}_{i}}{r_{ij}}=-
\frac{\sqrt{3}}{3} \sum_{k} (-1)^{k}
\sqrt{\frac{(k-1)(2k-1)}{2k+1}}
\frac{r_{<}^{k}}{r_{>}^{k+1}}\frac{r_{i}}{r_{j}}\times\left(\bi
{C}_{i}^{(k-1)}\times \left(\bi {C}_{j}^{(k)}\times
\bnabla_{\Omega_{j}}^{(1)}\right)^{(k-1)}\right)^{(1)}\nonumber\\
\!\!\!\!\!\!\!\!\!\!+ \frac{\sqrt{3}}{3} \sum_{k} (-1)^{k}
\sqrt{\frac{(k+1)(2k-1)}{2k+1}}
\frac{r_{<}^{k}}{r_{>}^{k+1}}\frac{r_{i}}{r_{j}}\times\left(\bi
{C}_{i}^{(k-1)}\times \left(\bi {C}_{j}^{(k)}\times
\bnabla_{\Omega_{j}}^{(1)}\right)^{(k)}\right)^{(1)}\nonumber\\
\!\!\!\!\!\!\!\!\!\!+ \frac{\sqrt{3}}{3} \sum_{k} (-1)^{k}
\sqrt{\frac{k(2k+3)}{2k+1}}
\frac{r_{<}^{k}}{r_{>}^{k+1}}\frac{r_{i}}{r_{j}}\times\left(\bi
{C}_{i}^{(k+1)}\times \left(\bi {C}_{j}^{(k)}\times
\bnabla_{\Omega_{j}}^{(1)}\right)^{(k)}\right)^{(1)}\nonumber\\
\!\!\!\!\!\!\!\!\!\!-\frac{\sqrt{3}}{3} \sum_{k} (-1)^{k}
\sqrt{\frac{(k+2)(2k+3)}{2k+1}}
\frac{r_{<}^{k}}{r_{>}^{k+1}}\frac{r_{i}}{r_{j}} \times\left(\bi
{C}_{i}^{(k+1)}\times \left(\bi {C}_{j}^{(k)}\times
\bnabla_{\Omega_{j}}^{(1)}\right)^{(k+1)}\right)^{(1)}\nonumber\\
\!\!\!\!\!\!\!\!\!\!+ \frac{\sqrt{3}}{3} \sum_{k} (-1)^{k}
\sqrt{\frac{k(k-1)(2k-1)}{(2k+1)^{2}}}
\frac{r_{<}^{k}}{r_{>}^{k+1}} r_{i}\frac{\partial}{\partial
r_{j}}\times\left(\bi {C}_{i}^{(k-1)}\times \bi {C}_{j}^{(k-1)}\right)^{(1)}\nonumber\\
\!\!\!\!\!\!\!\!\!\!- \frac{\sqrt{3}}{3} \sum_{k} (-1)^{k}
\sqrt{\frac{(k+1)(k+2)(2k+3)}{(2k+1)^{2}}}
\frac{r_{<}^{k}}{r_{>}^{k+1}} r_{i}\frac{\partial}{\partial
r_{j}}\times\left(\bi {C}_{i}^{(k+1)}\times \bi
{C}_{j}^{(k+1)}\right)^{(1)} \, . \quad \quad
\end{eqnarray}
When substituting $k \rightarrow k-1$ and $k \rightarrow k+1$ in the last two terms, respectively, 
the same angular contribution
$\left(\bi {C}_{i}^{(k)}\times \bi {C}_{j}^{(k)}\right)^{(1)}$ is obtained. 
After substituting $k \rightarrow k+1$ in the first two terms and summing 
over all electron pairs, one gets
\begin{eqnarray}
\label{E_ij_tens_1}
\fl\sum_{i<j}\bi {E}_{ij}=- \frac{\sqrt{3}}{3} \sum_{i\neq j}
\sum_{k} (-1)^{k}  \sqrt{(2k+1)(k+1)k}
\frac{r_{<}^{k}}{r_{>}^{k+1}}  \left(r_{<}^{2}
\frac{1}{2k+3}-r_{>}^{2}
 \frac{1}{2k-1}\right)
\frac{1}{r_{j}} \frac{\partial}{\partial r_{j}} \left(\bi
{C}_{i}^{(k)}\times \bi {C}_{j}^{(k)}\right)^{(1)}\nonumber\\
\!\!\!\!\!\!\!\!\!\!+\frac{\sqrt{3}}{3} \sum_{i\neq j} \sum_{k}
(-1)^{k} \sqrt{\frac{(2k+1)k}{2k+3}} \frac{r_{i}}{r_{j}}
\frac{r_{<}^{k+1}}{r_{>}^{k+2}}\times\left(\bi {C}_{i}^{(k)}\times
\left(\bi {C}_{j}^{(k+1)}\times
\bnabla_{\Omega_{j}}^{(1)}\right)^{(k)}\right)^{(1)}\nonumber\\
\!\!\!\!\!\!\!\!\!\!-\frac{\sqrt{3}}{3} \sum_{i\neq j} \sum_{k}
(-1)^{k} \sqrt{\frac{(2k+1)(k+2)}{2k+3}} \frac{r_{i}}{r_{j}}
\frac{r_{<}^{k+1}}{r_{>}^{k+2}}\times\left(\bi {C}_{i}^{(k)}\times
\left(\bi {C}_{j}^{(k+1)}\times
\bnabla_{\Omega_{j}}^{(1)}\right)^{(k+1)}\right)^{(1)}\nonumber\\
\!\!\!\!\!\!\!\!\!\!+\frac{\sqrt{3}}{3} \sum_{i\neq j} \sum_{k}
(-1)^{k} \sqrt{\frac{k(2k+3)}{2k+1}} \frac{r_{i}}{r_{j}}
\frac{r_{<}^{k}}{r_{>}^{k+1}}\times\left(\bi {C}_{i}^{(k+1)}\times
\left(\bi {C}_{j}^{(k)}\times
\bnabla_{\Omega_{j}}^{(1)}\right)^{(k)}\right)^{(1)}\nonumber\\
\!\!\!\!\!\!\!\!\!\!-\frac{\sqrt{3}}{3} \sum_{i\neq j} \sum_{k}
(-1)^{k} \sqrt{\frac{(k+2)(2k+3)}{2k+1}} \frac{r_{i}}{r_{j}}
\frac{r_{<}^{k}}{r_{>}^{k+1}} \times\left(\bi
{C}_{i}^{(k+1)}\times \left(\bi {C}_{j}^{(k)}\times
\bnabla_{\Omega_{j}}^{(1)}\right)^{(k+1)}\right)^{(1)}
\end{eqnarray}

\newpage

\subsubsection{Tensorial form of the two-body M1 operator} \ \\

The two-body magnetic dipole transition operator appearing in \eref{eq:M1_operator_split} has 
the following form
\begin{equation}
\label{sum_ij_Q_ij_1}
\sum_{i<j} \bi {Q}_{ij}=\mu_{B} \alpha^{2} \sum_{i<j} \left[\bi
{C}_{ij}+\frac{1}{2} (\bi {D}_{ij}-\bi {E}_{ij})\right]
\end{equation}
Replacing $\sum_{i<j} \bi {C}_{ij}$, $\sum_{i<j} \bi {D}_{ij}$ and $\sum_{i<j} \bi {E}_{ij}$ by 
their expressions
\eref{eq:sum_ij_C_ij_final}, \eref{D_ij_tens_2}, and \eref{E_ij_tens_1}, respectively,
we finally get the
irreducible tensorial form of the relativistic corrections to the magnetic transition operator
\begin{eqnarray}
\label{sum_ij_Q_ij_FINAL}
\fl\sum_{i<j} \bi {Q}_{ij}
=
-\frac{2}{3} \mu_{B} \alpha^{2}
\sum_{i\neq j} \sum_{k} (-1)^{k} \frac{r_{j}^{k}}{r_{i}^{k+1}}
\sqrt{2k+1} \; \varepsilon (r_{i}-r_{j}) \nonumber \\
 \times \Bigg[\left((\bi
{s}_{i}^{(1)}+\bi {s}_{j}^{(1)})\times\left(\bi
{C}_{i}^{(k)}\times \bi
{C}_{j}^{(k)}\right)^{(0)}\right)^{(1)} \nonumber \\
+ \sqrt{\frac{2k(k+1)}{(2k+3)(2k-1)}}\left((\bi {s}_{i}^{(1)}+\bi
{s}_{j}^{(1)})\times\left(\bi {C}_{i}^{(k)}\times \bi
{C}_{j}^{(k)}\right)^{(2)}\right)^{(1)} \nonumber \\
+\left(1-\frac{r_{j}^{2}}{r_{i}^{2}}\right)\sqrt{\frac{3(2k+5)(k+2)(k+1)}{4(2k+3)}}
 \left((\bi {s}_{i}^{(1)}+\bi {s}_{j}^{(1)})\times\left(\bi
{C}_{i}^{(k+2) } \times \bi
{C}_{j}^{(k)}\right)^{(2)}\right)^{(1)}\Bigg] \nonumber \\
\!\!\!\!\!\!\!\!\!\!+ \frac{\sqrt{3}}{6} \mu_{B} \alpha^{2}
\sum_{i\neq j} \sum_{k} (-1)^{k}  \frac{r_{i}}{r_{j}}
\frac{r_{<}^{k}}{r_{>}^{k+1}} \sqrt{2k+1} \nonumber \\
\times \Bigg[\sqrt{k(k+1)}\left(\frac{k+3}{2k+3} r_{<}^{2}+
\frac{k-2}{2k-1} r_{>}^{2}-r_{j}^{2}\right) \frac{1}{r_{i}}
\frac{\partial}{\partial r_{j}} \left(\bi
{C}_{i}^{(k)}\times \bi {C}_{j}^{(k)}\right)^{(1)} \nonumber \\
-\frac{r_{<}}{r_{>}}
\sqrt{\frac{k(k+3)^{2}}{2k+3}} \left(\bi
{C}_{i}^{(k)}\times\left(\bi {C}_{j}^{(k+1)}\times \bnabla
_{\Omega_{j}}^{(1)}\right)^{(k)}\right)^{(1)} \nonumber \\
+2\frac{r_{<}}{r_{>}}\sqrt{\frac{k+2}{2k+3}}
\left(\bi {C}_{i}^{(k)}\times\left(\bi {C}_{j}^{(k+1)}\times
\bnabla _{\Omega_{j}}^{(1)}\right)^{(k+1)}\right)^{(1)} \nonumber \\
-2\sqrt{\frac{k(2k+3)}{(2k+1)^{2}}} \left(\bi
{C}_{i}^{(k+1)}\times\left(\bi {C}_{j}^{(k)}\times \bnabla
_{\Omega_{j}}^{(1)}\right)^{(k)}\right)^{(1)} \nonumber \\
-
\sqrt{\frac{(k-1)^{2}(k+2)(2k+3)}{(2k+1)^{2}}} \left(\bi
{C}_{i}^{(k+1)}\times\left(\bi {C}_{j}^{(k)}\times \bnabla
_{\Omega_{j}}^{(1)}\right)^{(k+1)}\right)^{(1)}\Bigg] \; .
\end{eqnarray}

\section{Conclusion}

Systematic comparisons between different theoretical approaches and atomic structure codes 
are often used for assessing the reliability of the produced atomic data 
\cite{Paletal:05a,Froetal:06a,Hasetal:08a}. In this line, the Multiconfiguration 
Dirac-Hartree-Fock (MCDHF) and Multiconfiguration Hartree-Fock-Breit-Pauli (MCHF+BP) 
methods have been compared for transition probabilities in Fe IV of astrophysical 
interest \cite{FroRub:04a,Froetal:08b}. The authors of this comparison  
\cite{Froetal:08b} concluded that, although progress has been made since 
the pioneer work of Garstang \cite{Gar:58a}, agreement between MCHF+BP and MCDHF 
values for more transitions would be desirable. No doubt that the missing relativistic 
corrections considered in the present work should be systematically calculated within 
the first method for a definitive comparison. The tensorial form of the M1 transition 
operator derived in the present paper is the starting theoretical point for implementing 
the calculation of the relativistic corrections to the M1 transition probabilities in 
the atomic structure codes based on the Breit-Pauli approximation, such as RMATRX  
\cite{Beretal:95a}, CIV3 \cite{Hib:75a} or ATSP2K \cite{Froetal:07a} using Fano-Racah algebra 
\cite{Fanetal:63a,Fan:65a,GlaHib:76a,GlaHib:78a} or modern techniques combining second quantization 
and quasispin methods in coupled tensorial form \cite{Gaietal:97a,Gaietal:98a}.

\ack

M. Godefroid thanks the Communaut\'e fran\c{c}aise of Belgium (Action de Recherche Concert\'ee)
and the Belgian National Fund for Scientific Research (FRFC/IISN Convention) for financial support.
He is also grateful to Lidia Smentek, Jacques Breulet and Abdelatif Aboussa\"{\i}d for 
their earlier contributions to the present work. The authors acknowledge Elmar Tr\"{a}bert for helpful discussions.

\newpage
%
%%%%%%%%%%%%%%%%%%%%%%%%%%%%%%%%%%%%%%%%%%%%%%%%%%%%%%%%%%%%%%%%%%%%%%%%%%%%%%%%
%
%  APPENDIX
% %%%%%%%%%%%%%%%%%%%%%%%%%%%%%%%%%%%%%%%%%%%%%%%%%%%%%%%%%%%%%%%%%%%%%%%%%%%%%%%%
\appendix

\section{Intermediate calculations of the two-body contribution}
\label{app2}
 
The rank one tensors defined by
 \begin{eqnarray}
\fl \bi {F}_{ij}^{(1)} \equiv
-\frac{\sqrt{10}}{3}
\frac{r_{i}^{2}}{r_{>}^{2}-r_{<}^{2}} \sum_{k} (-1)^{k}
\frac{r_{<}^{k}}{r_{>}^{k+1}} (2k+1)^{3/2} \nonumber\\
\times \Bigg\{\left[(\bi {s}_{i}^{(1)}+\bi {s}_{j}^{(1)})\times
\bi {C}_{i}^{(2)}\right]^{(1)}\times \left[\bi {C}_{i}^{(k)}\times
\bi
{C}_{j}^{(k)}\right]^{(0)}\Bigg\}^{(1)} \; ,
\end{eqnarray}
\begin{eqnarray}
\fl \bi {H}_{ij}^{(1)} \equiv
+\frac{2}{3}\sqrt{15} \frac{r_{i}r_{j}}{r_{>}^{2}-r_{<}^{2}}
\sum_{k} (-1)^{k} \frac{r_{<}^{k}}{r_{>}^{k+1}} (2k+1)^{3/2}\nonumber\\
 \times \Bigg\{\left[(\bi {s}_{i}^{(1)}+\bi {s}_{j}^{(1)})\times
\left(\bi {C}_{i}^{(1)}\times \bi
{C}_{j}^{(1)}\right)^{(2)}\right]^{(1)}\times \left[\bi
{C}_{i}^{(k)}\times \bi {C}_{j}^{(k)}\right]^{(0)}\Bigg\}^{(1)} \; ,
\end{eqnarray} 
and appearing as the first and third terms of \eref{Y_ij_tens_1}, respectively, are 
transformed by first decoupling their spin and space parts and by using for the latter, 
the reduction formulae of tensor products involving four irreducible tensors given in \ref{B}. 
These contributions can then be rewritten as:

\begin{eqnarray}
\fl \bi {F}_{ij}^{(1)}= +\sqrt{\frac{3}{10}}
\sqrt{\frac{k(k-1)(2k-3)}{(2k+1)^{(2)}(2k-1)}} \left((\bi
{s}_{i}^{(1)}+\bi {s}_{j}^{(1)})\times \left(\bi
{C}_{j}^{(k)}\times \bi {C}_{i}^{(k-2)}\right)^{(2)}
\right)^{(1)}\nonumber\\
\nonumber\\
\!\!\!\!\!\!\!\!\!\!\!\!\!\!\!\!\!-\frac{1}{\sqrt{5}} \sqrt{\frac{k(k+1)}{(2k+3)(2k-1)}}
\left((\bi {s}_{i}^{(1)}+\bi {s}_{j}^{(1)})\times \left(\bi
{C}_{j}^{(k)}\times \bi {C}_{i}^{(k)}\right)^{(2)}
\right)^{(1)}\nonumber\\
\nonumber\\
\!\!\!\!\!\!\!\!\!\!\!\!\!\!\!\!\!+\sqrt{\frac{3}{10}}
\sqrt{\frac{(k+1)(k+2)(2k+5)}{(2k+3)(2k+1)^{2}}} \left((\bi
{s}_{i}^{(1)}+\bi {s}_{j}^{(1)})\times \left(\bi
{C}_{j}^{(k)}\times \bi
{C}_{i}^{(k+2)}\right)^{(2)}\right)^{(1)}\;\;\;
\end{eqnarray}

\vspace{0.5cm}

\begin{eqnarray}
\fl \bi {H}_{ij}^{(1)}=+\frac{1}{\sqrt{30}}
\sqrt{\frac{(2k-3)(2k-1)(k-1)k}{(2k+1)^{4}}}
%\nonumber\\ \quad\quad\quad\quad\quad\quad\quad\quad\times
\left((\bi
{s}_{i}^{(1)}+\bi {s}_{j}^{(1)})\times \left(\bi
{C}_{j}^{(k-1)}\times \bi {C}_{i}^{(k-1)}\right)^{(2)}
\right)^{(1)}\nonumber\\
\nonumber\\
\!\!\!\!\!\!\!\!\!\!\!\!\!\!\!\!-\frac{\sqrt{5}}{5}
\sqrt{\frac{(2k+3)(k+1)k(2k-1)}{(2k+1)^{4}}}
%\nonumber\\ \quad\quad\quad\quad\quad\quad\quad\quad\times
\left((\bi
{s}_{i}^{(1)}+\bi {s}_{j}^{(1)})\times \left(\bi
{C}_{j}^{(k-1)}\times \bi {C}_{i}^{(k+1)}\right)^{(2)}
\right)^{(1)}\nonumber\\
\nonumber\\
\!\!\!\!\!\!\!\!\!\!\!\!\!\!\!\!-\frac{\sqrt{5}}{5}
\sqrt{\frac{(2k+3)(k+1)k(2k-1)}{(2k+1)^{4}}}
%\nonumber\\ \quad\quad\quad\quad\quad\quad\quad\quad\times
\left((\bi
{s}_{i}^{(1)}+\bi {s}_{j}^{(1)})\times \left(\bi
{C}_{j}^{(k+1)}\times \bi {C}_{i}^{(k-1)}\right)^{(2)}
\right)^{(1)}\nonumber\\
\nonumber\\
\!\!\!\!\!\!\!\!\!\!\!\!\!\!\!\!+\frac{1}{\sqrt{30}}
\sqrt{\frac{(2k+3)(2k+5)(k+1)(k+2)}{(2k+1)^{4}}}
%\nonumber\\ \quad\quad\quad\quad\quad\quad\quad\quad\times
\left((\bi
{s}_{i}^{(1)}+\bi {s}_{j}^{(1)})\times \left(\bi
{C}_{j}^{(k+1)}\times \bi {C}_{i}^{(k+1)}\right)^{(2)}
\right)^{(1)}\nonumber\\
\end{eqnarray}

\section{Reduction of tensor products involving four irreducible tensors}\label{B}

\begin{eqnarray}
\label{CCCC(kk)0(11)1}
 \fl\left(\left(\bi {C}_{i}^{(k)}\times \bi {C}_{j}^{(k)}\right)^{(0)}
 \times
 \left(\bi {C}_{i}^{(1)}  \times \bi {C}_{j}^{(1)}\right)^{(1)}\right)^{(1)}
 =   \nonumber\\
\qquad \qquad \qquad + \frac{\sqrt{6}}{6}
\sqrt{\frac{(2k-1)(k-1)k}{(2k+1)^{3}}} \left(\bi
{C}_{j}^{(k-1)}\times \bi
{C}_{i}^{(k-1)}\right)^{(1)}\nonumber\\
\qquad \qquad \qquad - \frac{\sqrt{6}}{6}
\sqrt{\frac{(2k+3)(k+2)(k+1)}{(2k+1)^{3}}} \left(\bi
{C}_{j}^{(k+1)}\times \bi {C}_{i}^{(k+1)}\right)^{(1)}
\end{eqnarray}

%\vspace{0.5cm}

\begin{eqnarray}
\label{CCCC(kk)1(11)0}
\fl\left(\left(\bi {C}_{i}^{(k)}\times \bi {C}_{j}^{(k)}\right)^{(1)}\times \left(\bi {C}_{i}^{(1)}
\times
\bi {C}_{j}^{(1)}\right)^{(0)}\right)^{(1)}
 = \nonumber\\
\qquad \qquad \qquad
- \frac{\sqrt{3}}{3}
\sqrt{\frac{(2k-1)(k-1)(k+1)}{(2k+1)^{3}}} \left(\bi
{C}_{j}^{(k-1)}\times \bi
{C}_{i}^{(k-1)}\right)^{(1)}\nonumber\\
\qquad \qquad \qquad
- \frac{\sqrt{3}}{3}
\sqrt{\frac{(2k+3)(k+2)k}{(2k+1)^{3}}} \left(\bi
{C}_{j}^{(k+1)}\times \bi {C}_{i}^{(k+1)}\right)^{(1)}
\end{eqnarray}

%\vspace{0.5cm}

\begin{eqnarray}
\label{CCCnabla(kk)1(11)0}
 \fl\left(\left(\bi {C}_{i}^{(k)}\times \bi
{C}_{j}^{(k)}\right)^{(1)}\times \left(\bi {C}_{i}^{(1)}\times
\bnabla_{\Omega_{j}}^{(1)}\right)^{(0)}\right)^{(1)}=\nonumber\\
\qquad \qquad \;\;-\frac{\sqrt{3}}{3}
\sqrt{\frac{(2k-1)(k-1)(k+1)}{k(2k+1)^{2}}} \left(\bi
{C}_{i}^{(k-1)}\times \left(\bi {C}_{j}^{(k)}\times
\bnabla_{\Omega_{j}}^{(1)}\right)^{(k-1)}\right)^{(1)}\nonumber\\
\qquad \qquad \;\;+\frac{\sqrt{3}}{3}
\sqrt{\frac{(2k-1)}{k(2k+1)^{2}}} \left(\bi {C}_{i}^{(k-1)}\times
\left(\bi {C}_{j}^{(k)}\times
\bnabla_{\Omega_{j}}^{(1)}\right)^{(k)}\right)^{(1)}\nonumber\\
\qquad \qquad \;\;+\frac{\sqrt{3}}{3}
\sqrt{\frac{(2k+3)}{(k+1)(2k+1)^{2}}} \left(\bi
{C}_{i}^{(k+1)}\times \left(\bi {C}_{j}^{(k)}\times
\bnabla_{\Omega_{j}}^{(1)}\right)^{(k)}\right)^{(1)}\nonumber\\
\qquad \qquad \;\;+\frac{\sqrt{3}}{3}
\sqrt{\frac{(2k+3)(k+2)k}{(k+1)(2k+1)^{2}}} \left(\bi
{C}_{i}^{(k+1)}\times \left(\bi {C}_{j}^{(k)}\times
\bnabla_{\Omega_{j}}^{(1)}\right)^{(k+1)}\right)^{(1)}
\end{eqnarray}

In the first two formulaes, the tensorial operators 
$\bi {C}_{i}^{(k)}$, $\bi {C}_{i}^{(1)}$ et $\bi {C}_{j}^{(k)}$, $\bi {C}_{j}^{(1)}$ 
act in different spaces and all commute with each other.
To get expressions  \eref{CCCC(kk)0(11)1} and  \eref{CCCC(kk)1(11)0}, we have used 
the following transformation \cite{Varetal:88a}:

\begin{eqnarray}\label{CD}
\fl\left(\left(\bi {C}_{i}^{(k)}\times \bi {C}_{j}^{(k)}\right)^{(k_1)}
 \times
 \left(\bi {C}_{i}^{(1)}  \times \bi {C}_{j}^{(1)}\right)^{(k_2)}\right)^{(1)} = 
 \sum_{gh} (-1)^{g+h}\sqrt{5(2g+1)(2h+1)}
\left\{
\begin{array}{ccc}
k & k & k_1 \\
1 & 1& k_2 \\
g & h & 1
\end{array}
\right\} \nonumber\\
\times \left(\left(\bi {C}_{i}^{(k)}\times \bi {C}_{i}^{(1)}\right)^{(g)}
 \times
 \left(\bi {C}_{j}^{(k)}  \times \bi {C}_{j}^{(1)}\right)^{(h)} \right)^{(1)}
\end{eqnarray}
where the cases ($k_1=0/k_2=1$) and ($k_1=1/k_2=0$) correspond to  \eref{CCCC(kk)0(11)1} 
and  \eref{CCCC(kk)1(11)0}, respectively. The sum over $g$ and $h$ is limited to 
$g  = h = |k-1|$ and $g  = h = k+1$. By applying \eref{CK} to both tensorial products 
appearing in \eref{CD}, we get the right-hand side of equations \eref{CCCC(kk)0(11)1} 
and \eref{CCCC(kk)1(11)0}. To obtain \eref{CCCnabla(kk)1(11)0}, we also use the 
transformation \eref{CD} replacing  $\bi {C}_{j}^{(1)}$  by the operator 
$\bnabla_{\Omega_{j}}^{(1)}$ , keeping in the sum the ($h=k-1/g=k-1,\;k$) and 
($h=k+1/g=k,\;k+1$) contributions.

%%%%%%%%%%%%%%%%%%%%%%%%%%%%%%%%%%%%%%%%%%%%%%%%%%%%%%%%%%%%%%%%%%%%%%%%%%%%%%%%
%
%  BIBLIOGRAPHY
%
%%%%%%%%%%%%%%%%%%%%%%%%%%%%%%%%%%%%%%%%%%%%%%%%%%%%%%%%%%%%%%%%%%%%%%%%%%%%%%%%

%\bibliographystyle{unsrt}
%\bibliography{d:/mrg/atoms}
%\bibliography{I:/cpm-pc42/mrg/atoms}

\end{document}